\documentclass[twocolumn,showpacs,preprintnumbers,amsmath,amssymb]{revtex4}
\usepackage{graphicx}% Include figure files

\begin{document}
\preprint{OHSTPY-HEP-T-02-002}

\title{Statistical interpretation of Bekenstein entropy \\ for
systems with a stretched horizon}

\author{Oleg Lunin}
\author{Samir D. Mathur}%
       \affiliation{%
Department of Physics,
The Ohio State University,
Columbus, OH 43210
}%

\date{\today}% It is always \today, today,

\begin{abstract}
For the 2-charge extremal holes in string theory we show that the
Bekenstein entropy obtained from the area of the stretched horizon has a
statistical interpretation as a `coarse graining entropy': different
microstates give geometries that differ near $r=0$  and the
stretched horizon cuts off the metric at $r=b$ where these
geometries start to
differ.

\end{abstract}

\maketitle

\section{Introduction}

Black holes exhibit an intriguing thermodynamics. Gedanken experiments indicate
that a black hole has an entropy $S_{Bek}=\frac{A}{4G}$, where
$A$ is the area of the horizon and $G$ is the gravitational constant
\cite{bek}. The usual
principles of statistical mechanics then suggest that there should be
$e^S$ microstates
of the hole for given  macroscopic parameters of the hole. But the
metric of the hole
appears to be completely determined  by its macroscopic parameters
(a fact codified
in the conjecture  `black holes have no hair')  which means that the
gravitational theory is unable to exhibit the degeneracy required to
account for the
entropy.

String theory has in recent years made significant progress in black
hole physics. For extremal and
near extremal black holes we can count the  microstates for
a system of branes that carries the same energy and charges  as the
black hole, and
then we find an entropy $S_{micro}$ that equals the Bekenstein entropy
$S_{Bek}=\frac{A}{4G}$ of the corresponding hole \cite{svcm}. The
concept  of the AdS/CFT
correspondence \cite{mal} suggests that  $S_{micro}$  counts the
states of the hole in a
        field theory description {\it dual} to the gravitational description.

This still leaves the question: If we insist on using the
{\it gravitational} description of the
system, then are we supposed to find $e^S$  different states, and if
so, where in the
geometry would we find the differences that distinguish these
states from each
other?

We will look at the extremal two charge system in 4+1 noncompact
dimensions.  The microscopic entropy $S_{micro}$ of the corresponding
branes is nonzero.
But the horizon area of the (naively constructed) classical metric is 
zero, so that
$S_{Bek}$ appears to be zero.
In \cite{sen} the two-charge system was studied in 3+1 noncompact
directions.  Again $S_{micro}$ was
nonzero and the classical horizon area was zero.  But the curvature
diverged near $r=0$, so it was argued
that a `stretched horizon' should be placed at a location
$r=r_{stretch}$ where the curvature becomes
string scale. Further, the local temperature of the
Hawking radiation becomes of order the  Hagedorn temperature at
$r_{stretch}$.  Using the area of the
stretched horizon to write
$S_{Bek}=A_{stretch}/ 4G$ we find that
\begin{equation}
S_{Bek}\sim S_{micro}
\label{one}
\end{equation}
       in accord with the idea that $S_{Bek}$
is always a measure of the  entropy of the system.

In 4+1 noncompact dimensions the  2-charge extremal system has zero
temperature,
and so the local temperature is {\it zero} at all $r$.  But it was shown in
\cite{peet} that the
curvature again becomes string scale at some $r_{stretch}$, and (\ref{one}) is
satisfied by the corresponding stretched horizon.

While the large curvature at $r<r_{stretch}$ allows the possibility
that the geometry here may
be modified  by stringy corrections, it does not mean that such
corrections {\it must} occur. By a set of
dualities we can map the system to an extremal D1-D5 system, where
the near horizon
geometry is locally $AdS_3\times S^3\times T^4$.  Now the curvature
is {\it constant} at small $r$; further,
if the geometry were globally of this form then it would suffer no
quantum corrections whatever. In the
present case we have a discrete
identifications of points in the $AdS_3$ so it is not immediately
obvious what the corrections will
be.

      Thus the question arises: Is there a way to define the `stretched
horizon' so that we  get (\ref{one}) for all 2-charge systems related
by duality?
Further, does the construction of a   stretched horizon have the
interpretation of `coarse graining' over
different geometries, with the corresponding $S_{Bek}$ reflecting the
number of these geometries?

In this letter we argue that the physics of the stretched horizon
should not be thought of in terms of
quantum corrections to the naive metric. Instead,  we  note that
while the metrics corresponding to different
microstates of the matter system all look the same far from $r=0$,  close to
$r=0$ they are in
fact all {\it different}.  To coarse grain over such metrics we may therefore
truncate the
geometry at a radius $r\sim b$ where the geometries start differing from each
other.  Putting the stretched horizon at  $r_{stretch}=b$ we
find that the its area gives
$S_{Bek}\sim S_{micro}$.  Thus we find that
the Bekenstein entropy of this system  (computed from the area of the stretched
horizon) has a direct
interpretation in terms of a count of different metrics having the
same macroscopic
parameters. It is also evident that the result holds for all duality 
related 2-charge systems.

\section{The two-charge system}

We consider Type IIB string theory compactified on $T^4\times S^1$;
thus we will be
considering holes in 4+1 noncompact dimensions. The $T^4$ has volume
$(2\pi)^4 V$ and the
$S^1$ has length $2\pi R$ (we set $\alpha'=1$).

       Following \cite{sen} we consider a
fundamental string wrapped
$n_w$ times around the $S^1$, carrying momentum $P=\frac{n_p}{R}$
along the $S^1$.
These string states are BPS states, so that their mass is known for
all values of  the
string coupling $g$. This mass is $M= R n_w  + \frac{n_p}{ R}$.  The
entropy obtained from counting
string states
with this mass and charges is $S_{micro}\approx 2\sqrt{2} \pi \sqrt{n_pn_w}$.

We will refer to the above string states as the $FP$ system, where
$F$ stands for
fundamental string winding and $P$ stands for momentum. By a sequence of
$S$ and
$T$ dualities we can map this system to one having $n_1=n_p$ D1
branes wrapped on
the $S^1$, and $n_5=n_w$ D5 branes wrapped on $T^4\times S^1$. We
call this latter
system the D1-D5 system.

To achieve our desired goal of obtaining entropy by counting geometries we must
perform the following steps:

(a)\quad We must construct geometries corresponding to different
microstates of the
FP or D1-D5 system.

(b)\quad These geometries may have singularities at small $r$. In
that case we must
be satisfied that there is no further degeneracy associated to that
singularity, so that
the different geometries do indeed count different states of the system.

(c)\quad Finally we must locate the radius $r=b$ where the
geometries for generic microstates start to
differ from each other.
We must then compute the  area of a stretched horizon placed
at this location,
and compare the Bekenstein entropy obtained thereby with the count of
microstates.

\begin{figure}
\begin{tabular}{cc}
\begin{picture}(20,200)
\put(10,190){{\large a}}
\put(10,107){{\large b}}
\put(10,25){{\large c}}
\end{picture}&
\includegraphics{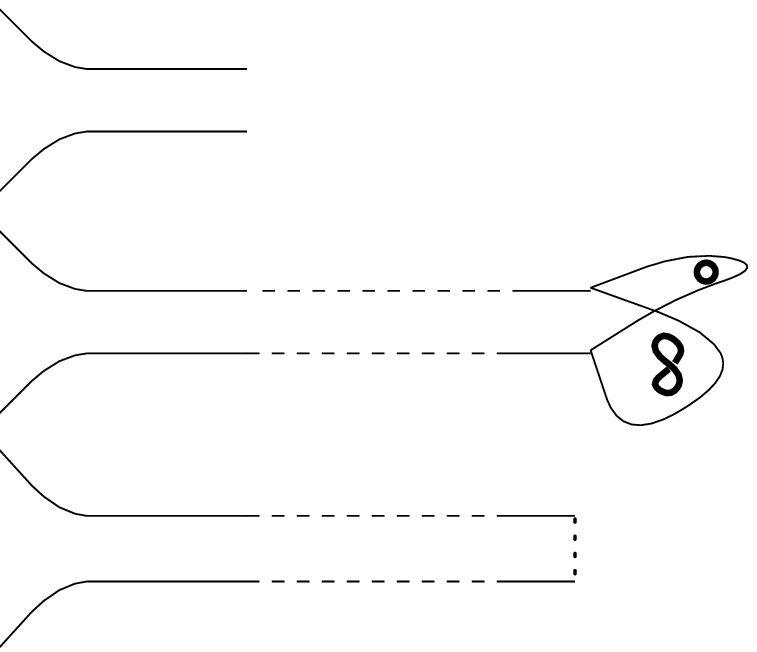}% Here is how to import EPS art
\end{tabular}
\caption{\label{fig1}(a) the geometry for $r\gg b$; (b) two different
possible ends to the throat;
(c) truncation by a stretched horizon.}
\end{figure}

\section{Geometries for different microstates}

To address (a) consider the FP system. The fundamental string wraps
the $S^1$ $n_w$
times before closing, so there are $n_w$ `strands' of the string at
any given value of
the coordinate $y$ parameterizing the $S^1$. The momentum excitation
gives traveling
waves along this string, creating vibrations in the 8 directions
transverse to the
string. In a generic vibration mode the different strands do not move
together --
they separate away from each other since the transverse displacement
8-vector $\vec
x$ satisfies the periodicity $\vec x(y+2\pi R n_w)=\vec x(y)$, rather
than $\vec
x(y+2\pi R )=\vec x(y)$.

The metric for a single string  carrying momentum  is known \cite{dh}, and the
metric  for the `multiwound string' having several strands can   be computed by
superposing  harmonic functions describing the individual strands. In
the classical
limit of large
$n_w$ one must then smooth out the strands into a continuous string
source. Such
metrics for the `multiwound string' were found in \cite{lm3} for a
special class of
vibration profiles, and extended in \cite{lm5} to generic vibration
profiles. For large
$n_w$ the vibration profile for a generic mode changes by a very
small amount when
$y\rightarrow y+2\pi R$, so the solutions in the classical limit are
found to be $y$
independent and a T-duality may be performed along the $S^1$.  Using
this and other
dualities the solutions are mapped to geometries for the D1-D5 system
\cite{lm5}.

The metric for a general D1--D5 bound state in then found to have the form
\begin{eqnarray}\label{fluctGeom}
ds_E^2&=&\frac{1}{\sqrt{g_1g_5}}\left[-(dt-A_i dx^i)^2+(dy+B_i dx^i)^2\right]
\nonumber\\
&+&\sqrt{g_1g_5}\sum_{i=1}^4dx_idx_i,
\end{eqnarray}
where the functions $g_1$, $g_5$ and $A_i$ are given in terms of an
arbitrary function
${\vec F}(v)$
\begin{eqnarray}\label{fluctCoeffG}
g_1({\vec x})&=&1+\frac{Q_5}{L}\int_0^L\frac{|{\dot{\vec F}}(v)|^2dv}{
|\vec x-\vec F(v)|^2},\\
g_5({\vec x})&=&1+\frac{Q_5}{L}\int_0^L\frac{dv}{
|\vec x-\vec F(v)|^2},\\
\label{fluctCoeff}
A_i({\vec x})&=&-\frac{Q_5}{L}\int_0^L\frac{{\dot F}_i(v)dv}{
|\vec x-\vec F(v)|^2}
\end{eqnarray}
We will not give a form of the field $B_i$ here, since it will not be
used. In the dual FP system $\vec
F(v)$ yields the transverse displacement of the vibrating string as a
function of the null coordinate $v=t+y$.

If $|\vec F(v)|<b$, then for $r\gg b$ the metric has the form \cite{lm5}
\begin{eqnarray}\label{NonRotGeom}
ds_E^2&=&\frac{1}{\sqrt{g_1g_5}}\left[-dt^2+dy^2\right]
+\sqrt{g_1g_5}\left(dr^2+r^2d\Omega_3^2\right)\nonumber\\
&&g_1(r)=1+\frac{Q_1}{r^2},\qquad  g_5(r)=1+\frac{Q_5}{r^2},
%&& Q_1=\frac{Q_5}{L}\int_0^L|{\dot{\vec F}}(v)|^2dv
\end{eqnarray}
The charges are related by $Q_1=\frac{Q_5}{L}\int_0^L|{\dot{\vec F}}(v)|^2dv$.

The location $r\sim b$ will turn out such that  a  massless particle
falling radially down the
throat' takes a time
$\Delta t \sim R\sqrt{n_1n_5}$ to reach  $r \sim b$. For fixed
classical charges $Q_1, Q_5$ this time goes to
{\it infinity} as $\hbar\rightarrow 0$. Thus if we look upto any
`classical' distance down the throat then
all the geometries
look the same (eqn. (\ref{NonRotGeom})), so that we see `no hair'.

But around $r\sim b$ the different metrics start to differ from each
other. Further,
they all have an `end', which contains a mild singularity along a
certain curve -- the
shape of this curve depends on the chosen microstate. These different
geometries are
schematically sketched in Fig 1(b).  Each  function $\vec F(v)$ gives
one extremal   D1-D5
geometry, which means that for each classical profile of the
oscillating string in the FP system there is a classical
D1-D5 geometry. We take this to imply that if we quantize the metric
of the D1-D5 system we will get one state
of the throat  for each quantum state of the string in the FP system.
This gives us the count of
microstates arising from the different possible geometries of the
D1-D5 system with given total charges:
$S_{micro}=2\sqrt{2}\pi\sqrt{n_1n_5}$.

We now address the requirement (b). In \cite{lm4} the propagation of a scalar
was studied in the geometries corresponding to different states of
the D1-D5 system. The CFT dual of these
geometries can be described through an `effective string' which
carries massless bosonic and fermionic vibration
modes as its low energy dynamics. For a class of geometries where the
wavepacket traveled without distortion it
was found that the travel time for the wavepacket in the `throat'
{\it exactly} equaled the time in the dual CFT for
the vibration modes to travel around the effective string:
\begin{equation}
\Delta t_{CFT}=\Delta t_{SUGRA}
\end{equation}
The wavepacket could not `enter' into the singularity and spend
additional time residing at the singularity.
It was then argued that for the generic geometry (where the
singularity was a curve with more
complicated shape) a similar result held: the singularity was mild
enough that the wavepacket reflected
from the singular curve rather than enter into the singular region. Thus we
conclude that the `throat' of the geometry is indeed a dual
representation of the D1-D5 CFT with no additional
degrees of freedom being associated to the singularity at the end of
the throat.
\begin{figure}
\includegraphics{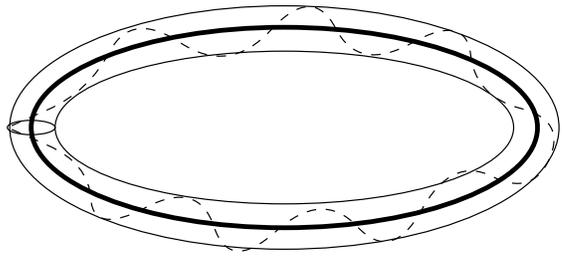}
%\epsfysize=1.2in \epsffile{PRLTor.eps}
% Here is how to import EPS art
\caption{\label{fig2}A typical singular curve (dashed line) and
stretched horizon (torus surface) for $J\gg \sqrt{n_1n_5}$.}
\end{figure}

We now perform step (c). We let the geometry be (\ref{NonRotGeom}) for $r>b$
and put a `stretched horizon' at the location $r=b$
where the generic geometries start to depart from the form
(\ref{NonRotGeom}) (Fig \ref{fig1}(c)).
The parameter $b$ will be determined shortly.
The area of the stretched horizon (in the 6-D geometry) is then
\begin{equation}
A=\int_{r=b} r^3\sqrt{g_1g_5}dy d\Omega_3\approx 4\pi^3Rb\sqrt{Q_1Q_5}.
\end{equation}
(We have assumed $b\ll Q_1, Q_5$ which will be well satisfied in the
classical limit.) To determine $b$ we start with the
FP system, and note that the `multiwound string' has a total length
$L_T=2\pi R n_w$ and carries total
momentum $2\pi n_pn_w/L_T$.  A statistical analysis shows that the
mean wavelength of vibration is
$\lambda\sim \bar\gamma L_T$ with $\bar \gamma\sim 1/\sqrt{n_1n_5}$.
If we take a string carrying vibrations with $\lambda\sim \bar\gamma L_T$, then
we find upon dualizing to D1--D5 \cite{lm3}:
\begin{equation}
b\sim \frac{\sqrt{Q_1Q_5}\bar\gamma}{R}\sim \frac{g}{\sqrt{V}R}
\end{equation}
(In this calculation we need to filter out the low energy tail of the energy
distribution; the details of this will be discussed elsewhere)
We can
now find the entropy of the stretched horizon
($G^{(6)} =\frac{8\pi^6 g^2}{(2\pi)^4 V}$)
\begin{equation}
S_{Bek}\equiv \frac{A^{(6)}}{ G^{(6)}}=\frac{A^{(5)}}{G^{(5)}}\sim
\sqrt{n_1n_5}
\end{equation}
which agrees with $S_{micro}\sim 2\pi\sqrt{2}\sqrt{n_1n_5}$.

We now extend the above analysis to the case where the 2-charge
system also carries angular momentum. Let the
angular momentum be $J\hbar$ in in the $x_1-x_2$ plane (these are two
of the four
noncompact directions). An analysis of the microstate (in the FP language)
gives
$S_{micro}=2\pi\sqrt{2}\sqrt{n_1n_5-J}$.  We then
find the  geometries  for different FP microstates with angular
momentum $J$ and dualize these to geometries for the D1-D5 system. The metrics
are given by (\ref{fluctGeom})--(\ref{fluctCoeff}) with
\begin{equation}\label{RotProfile}
{\vec F}(v)=a~{\vec e}_1\cos\frac{2\pi v}{ L}+a~{\vec e}_2\sin\frac{2\pi
v}{ L}+{\vec X}(v)
\end{equation}
The singularity now lies close to  a
circle of radius
\begin{equation}
a=\frac{g}{\sqrt{V}R}\sqrt{J}
\end{equation}
in the $x_1-x_2$ plane. The generic geometries differ from each other
only in a tube around this circle (each state is specified by its own
fluctuation profile ${\vec X}(v)$: $|{\vec X}(v)|<b$), so the stretched
horizon has the shape
of a `doughnut' in the noncompact
space $x_1, x_2, x_3, x_4$ (Fig \ref{fig2}). Again performing  a
statistical analysis of the vibrations in the FP system, we find
$b\sim \frac{g}{\sqrt{V}R}$.
All geometries described by the profile (\ref{RotProfile}) look similar
outside the `doughnut', and the coefficients of the generic metric
(\ref{fluctGeom}) can be found using (\ref{fluctCoeffG})--(\ref{fluctCoeff}):
\begin{eqnarray}
&&g_1=1+\frac{Q_1}{f_0},\qquad g_5=1+\frac{Q_5}{f_0},\\
&&A_i dx^i=\sqrt{\frac{J}{n_1n_5}}
\frac{2\sqrt{Q_1Q_5}a}{f_0({\vec x}\cdot {\vec x}+a^2+f_0)}
(x_2dx_1-x_1dx_2),\nonumber\\
&&f_0=\left[({\vec x}\cdot {\vec x})^2+2a^2(x_3^2+x_4^2-x_1^2-x_2^2)+
a^4\right]^{1/2}
\end{eqnarray}
The area of the stretched horizon (computed at fixed $t$) for this metric
finally gives
\begin{eqnarray}
S_{Bek}&\equiv&\frac{A^{(6)}}{4G^{(6)}}=
\frac{VR}{\pi g^2}\int d^3\Sigma~ (g_1g_5-A_iA_i)^{1/2}\nonumber\\
&\sim& \sqrt{n_1n_5-J}~\sim~ S_{micro}
\end{eqnarray}

\section{Discussion}
\begin{figure}
\includegraphics{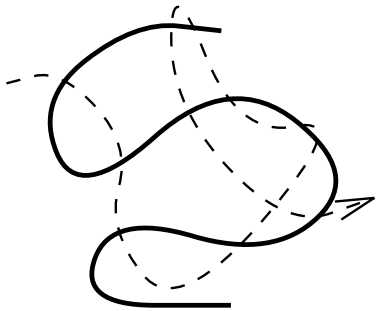}
%\epsfysize=1.2in \epsffile{PRLTor.eps}
% Here is how to import EPS art
\caption{\label{fig3} Typical geodesic (dashed line) near the
singular curve (solid line).}
\end{figure}
Our conventional understanding of entropy is based on coarse graining
over a large
number of microstates. Black hole entropy has proved puzzling since
black holes seem to
have `no hair' and the entropy $S_{Bek}$ is instead given by the
horizon area. We have
seen above that at least for the 2-charge system (which has a
stretched horizon rather
than a classical horizon) we do indeed have a complete set of `hair', that the
appearance of the stretched horizon can be regarded as a `coarse
graining' since it
truncates the geometries where they start to differ from each other,
and that the area of
such a stretched horizon gives a $S_{Bek}$ which is of order the entropy
$S_{micro}$ found by
actually counting the different allowed geometries.

A crucial ingredient in the above result was the fact that the D1-D5
bound states had a
nonzero size; this caused the `throat' to end at some point before
reaching $r=0$, with a
metric near the end that reflected the choice of microstate. This
nonzero size could itself
be traced back, through duality to the FP system, to the fact that a
string (F) carrying
momentum (P) must spread over a certain transverse region in order to carry the
momentum. Thus the nonzero size and the consequent `hair' are a very
basic feature of
the structure of 2-charge states.

Should we regard the stretched horizon of the 2-charge system as a
black hole horizon?
We performed a detailed investigation (to be presented elsewhere) of
the trajectory of a
massless particle that falls into the region $r<b$. For a generic
microstate the singular
curve of the metric (\ref{fluctGeom}) is a complicated `random curve' in the
transverse coordinates
$x^1\dots x^4$. A null geodesic gets deflected through an angle of
order unity on passing
near any point on this curve, and as a consequence the particle stays
trapped in the
region $r<b$ for times that go to infinity in the classical limit
(Fig \ref{fig3}). If
we choose to describe the
particle by its wavefunction instead then we expect  large trapping
times due to the
presence of approximately localized wavefunctions in the `random 
potential' arising
from the metric at
$r<b$.  Thus it appears that we should regard the stretched
horizon as a horizon, with the time delay in emerging from the horizon
being due to
`trapping' in the hair describing the microstate.

A similar picture may emerge for the
D1-D5-momentum hole (which
has a classical horizon area). 
The size of the 3-charge bound state
at weak coupling was argued to have the same algebraic expression as 
the horizon radius \cite{mathur}.
The metrics for all microstates are
the same upto a `classical
distance' down the throat, but the throats may end further down, with the
geometry near the end characterizing the
microstate. In view of the
fact that the momentum charge generates vibrations within the $T^4$ we
expect that in
this case both the $S^3$ and the $T^4$ will necessarily be
deformed near the end of the
throat.

We thank E. Braaten, J. Ho and Ashoke Sen for
helpful discussions. This work was supported in part by DOE grant
DE-FG02-91ER40690.

\bibliography{apssamp}% Produces the bibliography via BibTeX.

\end{document}